# Enhanced Performance of Short-Channel Carbon Nanotube Field-Effect Transistors Due to Gate-Modulated Electrical Contacts

*Aron W. Cummings and François Léonard*

Sandia National Laboratories, MS9161, Livermore, California 94551, USA

awcummi@sandia.gov

We use numerical simulations to analyze recent experimental measurements of short-channel carbon nanotube field-effect transistors with palladium contacts. We show that the gate strongly modulates the contact properties, an effect that is distinct from that observed in Schottky barrier carbon nanotube transistors. This modulation of the contacts by the gate allows for the realization of superior subthreshold swings for short channels, and improved scaling behavior. These results further elucidate the behavior of carbon nanotube-metal contacts, and should be useful in the optimization of high-performance carbon nanotube electronics.

Due to their unique electrical properties, carbon nanotubes (CNTs) have attracted a great deal of interest for their potential in next-generation nanoelectronics.[1,2] While individual CNTs can exhibit favorable electronic properties, it is often the CNT/metal contacts that govern the behavior and performance of CNT devices.[3,4] Thus, it is important to develop a fundamental understanding of contacts to CNTs in order to fully realize the potential of CNT devices. While work on CNT/metal contacts has addressed the issues of band alignment,[5-8] charge injection,[9-11] and structural properties,[12-14] many questions still remain. Recent experimental work[15,16] has provided new insight by demonstrating that the nanotube/palladium (Pd) contact resistance depends on the contact length, and that appropriate control of the contacts allows for the realization of high-performance short-channel CNT field-effect transistors (FETs) with subthreshold swings that surpass those expected from conventional scaling theory. This last result is particularly important not only for technology, but also because it suggests that new paradigms govern the properties of these nanoscale transistors. For example, it has been suggested that modulation of the contacts by the gate, a phenomenon not usually observed in conventional transistors, could lead to such behavior.[16] The gate modulation of contacts to graphene nanoribbons has also been studied recently.[17]

In this paper, we use numerical simulations to study these recent experimental measurements and explicitly demonstrate that the superior scaling behavior is due to a strong modulation of the contacts by the gate. This results not only in modulation of the band alignment at the contact, but also leads to a novel phenomenon where the subthreshold swing is dominated by gate control of the near-contact region in the channel. This gives rise to subthreshold swings for short-channel devices that are below what is predicted by standard theory, allowing for improved performance. In addition, we show that field enhancement at the dielectric/CNT interface plays an important role in augmenting the impact of the gate on the contacts.

The CNT FET to be simulated is shown in Figure 1. For this work, we consider a (16,0) nanotube with a diameter ($d_{CNT}$) of 1.2 nm, which matches the average size of the CNTs in Ref. 15. The dielectric is $HfO_2$, the oxide thickness ($t_{ox}$) is 10 nm, the height of the source/drain contacts ($t_c$) is 20 nm, the CNT is separated from the oxide or the metal by 0.3 nm, and there are 100 nm of vacuum above the source/drain contacts ($t_{vac}$). We use a dielectric constant of 20 for the $HfO_2$ layer, yielding an equivalent oxide thickness of 2 nm. The channel length ($L_{ch}$) and the contact length ($L_C$) are both variable. Figures 1a and 1b differ in the geometry of the contact. In Figure 1a, there is metal both above and below the nanotube, as a model for a CNT completely embedded in metal. In Figure 1b, we consider a contact where the metal only sits on top of the CNT. The type of metal is defined by the difference between its work function and that of the CNT, $\Delta\phi = \phi_{CNT} - \phi_{metal}$. The value of $\Delta\phi$ then determines the potential in the contacts, assuming the reference is at the CNT mid-gap. In all cases, we assume a temperature of 300 K.

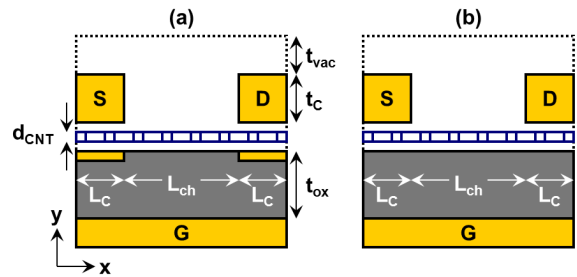

Figure 1. Schematic of a carbon nanotube field-effect transistor. In part (a) the source and drain metals are above and below the nanotube (embedded contact), while in part (b) the metal only sits on top of the nanotube (top contact).

To determine the transport properties of the FET, we use a self-consistent non-equilibrium Green's function (NEGF) approach.[18] The first step is a self-consistent calculation of the charge and potential within the FET for a given gate voltage ($V_G$). The potential is obtained from the charge through a 3D solution of Poisson's equation, $\nabla \cdot (\varepsilon \nabla V) = -\rho$, where $\rho$ is the charge density, $V$ is the electrostatic potential, and $\varepsilon$ is the spatially-dependent dielectric constant. To calculate the charge due to a given potential, we use a tight-binding model for the CNT, and use the NEGF approach to determine the electron correlation function. The coupling between the CNT and the metal contacts is described by the parameter $\Delta$, which is a measure of the CNT-metal hybridization.[10,11] In this representation, $\Delta = 0$ in the channel and $\Delta > 0$ in the contacts. Due to less metal coverage of the CNT, we expect that the average value of $\Delta$ will be smaller for the top contact than for the embedded contact. For the contact lengths that we consider, this would result in an increased contact resistance for the top contact, which would reduce the on current of the FET. However, this will have no effect on the gate modulation of the top contact or the improved subthreshold performance. Therefore, we have chosen $\Delta$ to be the same for both contacts in order to treat them on equal footing and isolate the effect of their differing geometries. Once the charge and the potential have been determined self-consistently, the ballistic zero-bias conductance, $G$, can be calculated using the NEGF formalism. Finally, the small-bias current through the device is given by $I_D = G \cdot V_{DS}$, where $V_{DS}$ is the source-drain bias.

RESULTS AND DISCUSSION

A central result of the experimental work of Ref. 15 is a strong dependence of contact resistance on contact length. Thus, before simulating the CNT transistor characteristics, we parameterized our contact model by fitting to the contact resistance data of Ref. 15. To calculate the contact resistance, we assumed perfectly flat bands along the CNT. In this case, the channel resistance vanishes and the contact resistance is given by $2R_C = 1/G$, where $G$ is the zero-bias conductance. By adjusting two independent parameters, $\Delta$ and $\Delta\phi$, we obtain an excellent match to the experimental data. This is shown in Figure 2, where we plot the contact resistance as a function of the contact length. The symbols represent the experimental results and the solid line shows our fit to the experimental data, assuming $\Delta = 2.5$ meV and $\Delta\phi = -0.7$ eV. The work function difference is reasonable for a CNT-Pd contact,[12,19] assuming $\phi_{CNT} = 4.5$ eV and $\phi_{Pd} = 5.2$ eV. Furthermore, the self-consistent band alignment puts the Pd Fermi level 24 meV below the CNT valence band edge for the embedded contact, which indicates that the contact is ohmic in the on-state. For the top contact, the Fermi level varies from 19 to 59 meV below the valence band edge in the on-state, for $V_G = 0$ to -0.8 V.

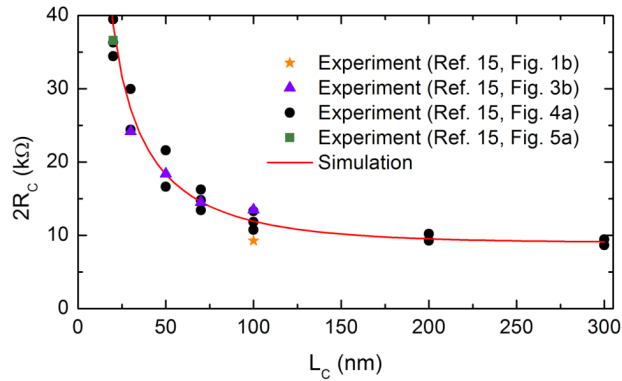

Figure 2. Contact resistance as a function of contact length. The symbols represent the experimental values extracted from Ref. 15. The solid line is the fit assuming a CNT-metal coupling strength of $\Delta = 2.5$ meV and a work function difference of $\Delta\phi = -0.7$ eV.

Using these values of $\Delta$ and $\Delta\phi$, we then calculated the transfer characteristics of the CNT FETs for channel and contact lengths that match the experimental devices. The results are shown in Figure 3, where the experimental data is given by the symbols and the theoretical data is given by the solid lines (the experimental data has been shifted horizontally to line up with the theoretical data; this discrepancy could be due to trapped charges in the oxide, interface charge between the gate and the oxide, or other effects). The top row of Figure 3 shows the results for $L_{ch}$ = 40 nm, the middle row is for $L_{ch}$ = 20 nm, and the bottom row is for $L_{ch}$ = 15 nm. The left column shows the simulation results for embedded contacts (see Figure 1a), while the right column is for top contacts (see Figure 1b). The experimental data is the same for both columns.

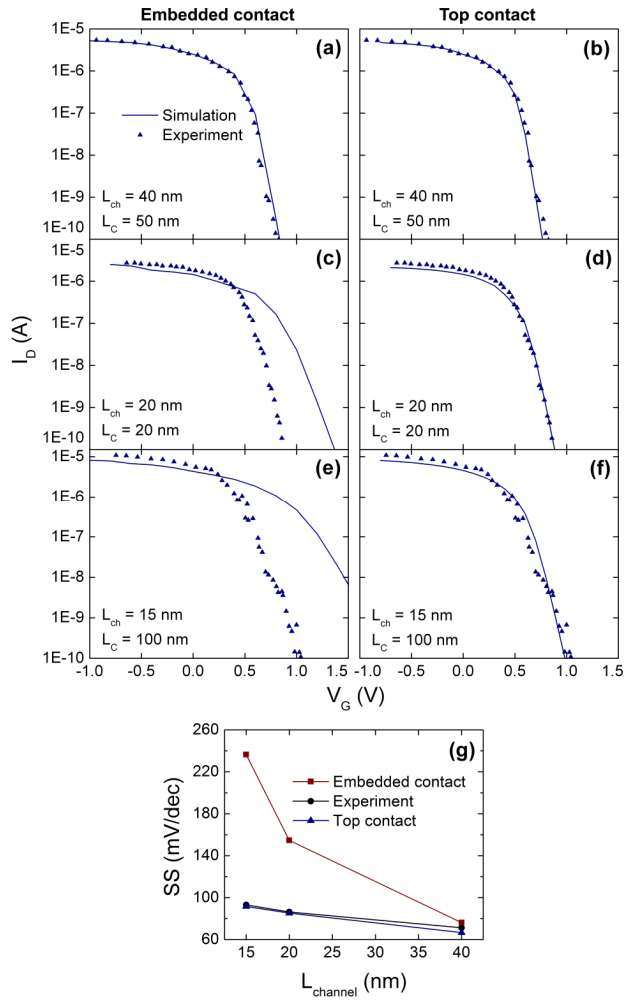

Figure 3. Panels (a)-(f) show the current vs. gate voltage for a variety of CNT FETs. The top, middle, and bottom rows are for $L_{ch}$ = 40, 20, and 15 nm, respectively. The left (right) column is the case for embedded (top) contacts. The symbols represent experimental results from Ref. 15, and the solid lines represent the results from numerical simulations. Panel (g) shows the subthreshold swing as a function of channel length. The black circles are for the experimental devices, the red squares are for embedded contacts, and the blue triangles are for top contacts.

     An important feature of the experimental data is the extremely good scaling of the transistor characteristics as the channel length is reduced. Indeed, comparing the experimental data for the channel lengths of 40, 20, and 15 nm in Figure 3, one can see that the subthreshold swing is essentially unchanged as the channel length is scaled down. While the thin $HfO_2$ dielectric provides good control over the FET channel, our simulations indicate that this by itself is not sufficient to explain the good subthreshold behavior. This can be seen by comparing the left and right columns of Figure 3. The left column shows the simulation results for the embedded contacts. In this case, the theoretical subthreshold swing is much larger than the experimental value for small channel lengths, and we see a poor fit to the experimental results. However, when we remove the metal below the CNT, the subthreshold swing is significantly reduced for the short-channel devices and we obtain excellent agreement with the experimental data, as shown in the right column of Figure 3. These results are summarized in Figure 3g, where we plot the subthreshold swing as a function of channel length for the experimental devices (black circles), for embedded contacts (red squares), and for top contacts (blue triangles). Here, one can see that the experimental subthreshold swing is well matched by the top contact geometry and short-channel effects are strongly mitigated in contrast to the embedded contacts.

Thus, the geometry of the contact plays a crucial role in determining device performance and scaling, and the improved behavior upon removing the bottom metal suggests an influence of the gate on the contact properties.

To understand this effect, in Figure 4 we plot the energy band profile of the CNT for $L_{ch}$ = 15 nm. The dashed line indicates the Fermi level, and the solid lines indicate the conduction and valence band edges. Figures 4a and 4b show the band edges for $V_G$ = -0.8 V, for embedded and top contacts, respectively. At this gate voltage, the contacts are ohmic and the FET is in the on-state, and the presence of the bottom metal has little effect on the on-state performance. However, the situation in the off-state is quite different, as shown in Figs. 4c and 4d. There, the embedded contact is still ohmic, and in the channel the gate has pulled the valence band edge below the Fermi level, slightly turning off the transistor. Without the bottom metal (Figure 4d), the gate has pulled down the bands in the contacts by ~84 meV, giving a Schottky barrier of ~60 meV. The introduction of this gate-dependent Schottky barrier leads to a faster turning off of the FET compared to the embedded contact geometry. However, while this 84 meV variation is large and significantly improves the subthreshold swing, it only partially explains the impact of the gate. Indeed, one can also see in Figure 4 that the hole barrier created by the gate is much larger for the case of the top contact (281 meV) compared to the embedded contact (113 meV). Thus, the gate is more effective at modulating the channel potential in the case of the top contact. It might appear that the extra screening due to the bottom metal explains this phenomenon, but we have verified that this is not the case.

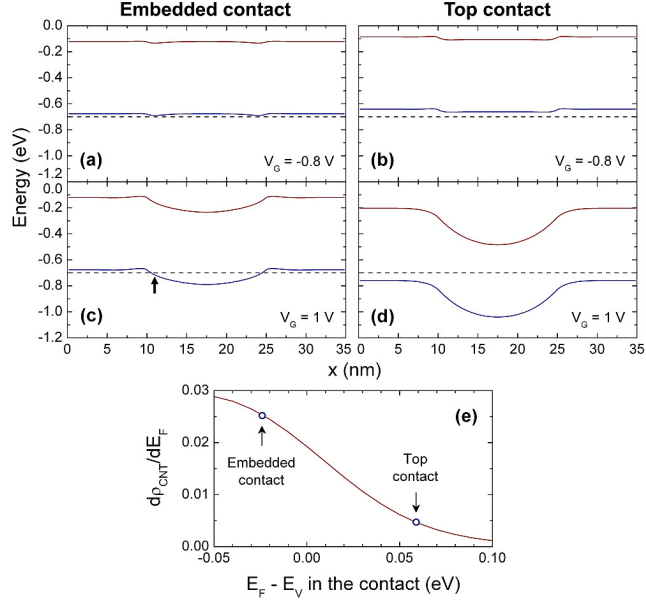

figure 4. Panels (a)-(d) show the band-edge profile of the CNT FET for $L_{ch}$ = 15 nm. The left (right) column is for an embedded (top) contact. The top (bottom) row shows the case for $V_G$ = -0.8 V (1 V). Panel (e) shows the differential charge with respect to the Fermi energy at the location of the arrow in panel (c).

Instead, the effect is a direct consequence of the gate modulation of the contact and can be understood as follows: to establish a particular potential in the middle of the channel, the gate has to create an appropriate band-bending in a region between the contact and the middle of the channel. The effectiveness of the gate in modulating the charge in this region depends on how strongly the charge on the CNT varies with the potential there. To illustrate this, in Figure 4e we plot the differential charge on the CNT, $d\rho_{CNT}/dE_F$, as a function of the Fermi energy, calculated at a position indicated by the arrow in Figure 4c. When the Fermi energy is near the valence band edge, as for ohmic contacts, the differential charge is large and the gate has little control over the near-contact region because the bands are effectively pinned by the large density of states. However, when the contacts are Schottky the differential charge at the Fermi level is small and the gate has much more control over the bands in the near-contact region, which results in a larger barrier in the middle of the channel.

In our simulations, we used a geometry with top and bottom planar metals to illustrate the impact of the gate on the contacts. However, experimentally it is expected that the metal will surround the CNT, except at the bottom where the CNT sits on the oxide. It may appear surprising then that the gate fields are able to penetrate the tiny cavity where the CNT sits. To address this point, in Figure 5 we consider the electrostatics of the cavity from an end-on view. In this figure, the cavity is a rectangular hole of width $W_C$ in the contact metal (dashed white lines), which sits on top of the 10-nm oxide, and we calculate the electrostatic potential by solving Laplace's equation for $V_G = 1$ V. In the limit $W_C \to \infty$, and for a dielectric extending to the top contact, the electrostatic potential drops linearly from the gate to the contact. In this case, the potential at the center of the CNT is roughly $\left[ \left( R_{CNT} + 0.3 \text{ nm} \right) / t_{ox} \right] \cdot V_G$, which is about 10% of $V_G$. Thus, a 1V change of the gate voltage can shift the band alignment at the nanotube/metal contact by 100 meV despite the proximity of the CNT to the contact metal. When the oxide is removed from the CNT region, the potential is no longer linear due to the dielectric/vacuum interface, and the potential on the CNT increases to more than 40% of $V_G$, illustrated by the dashed line in the inset of Figure 5.

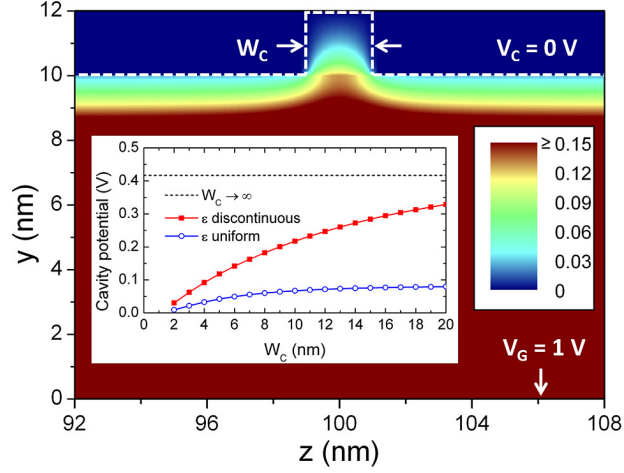

Figure 5. Electrostatics of the top contact geometry. The dashed lines show the edge of the contact metal, and the color scale indicates the electrostatic potential for the given gate and contact voltages. The inset shows the potential at the middle of the cavity as a function of the cavity width, $W_C$. The filled squares are for a vacuum in the cavity, and the open circles are for an oxide-filled cavity.

To assess whether decreasing the cavity size can prevent field penetration, we calculated the potential in the center of the cavity as a function of $W_C$. The solid squares in the inset of Figure 5 show the case for $HfO_2$ with a vacuum in the cavity, and the open circles show the case when the cavity is also filled with $HfO_2$. We can see that as $W_C$ decreases, so does the cavity potential. This is understandable, since decreasing $W_C$ increases the screening of the gate by the contact metal. Nevertheless, even as the cavity size decreases to 2 nm, the potential in the middle of the vacuum cavity is still 30 meV, sufficient to change the band alignment at the contact from ohmic to Schottky. This strong penetration of the gate fields into the cavity is illustrated in the color plot of Figure 5. Note that if the cavity is filled with the same material as the gate oxide (open circles in the inset of Figure 5), the potential is much smaller. Thus, it is the discontinuity of the dielectric constant at the dielectric/vacuum interface that allows for the high degree of gate control over the contact.

The results presented above demonstrate a significant improvement of the subthreshold performance when we consider the gate modulation of the contacts. Looking forward, this raises an interesting question concerning the ultimate scaling limits of these devices; how close can these devices come to the room-temperature limit for subthreshold swing? To examine this question, in Figure 6 we plot the subthreshold swing as a function of oxide thickness for three different cases: an embedded contact with a channel length of 15 nm (blue triangles), and a top contact with channel lengths of 15 nm (black squares) and 40 nm (red circles). For the top contact with a 40-nm channel length, the subthreshold swing is 62 mV/dec for an oxide thickness of 1 nm, which approaches the room-temperature limit of 60 mV/dec.[20] More interestingly, we also see that the subthreshold swing for the top contact changes relatively little as the oxide thickness decreases. This highlights the strong impact that gate modulation of the contacts has on device performance. It also suggests that when the gate modulation effect is present, scaling the oxide thickness below a critical value is not necessary for the realization of high-performance CNT FETs.

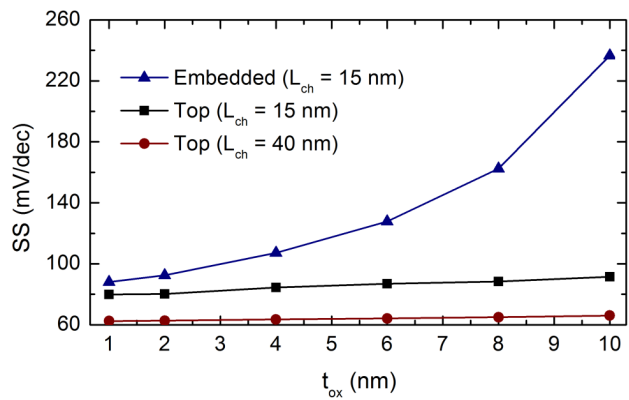

Figure 6. Subthreshold swing as a function of oxide thickness for three different CNT FETs. Blue triangles are for an embedded contact with a channel length of 15 nm, and the black squares

(red circles) are for a top contact with a channel length of 15 nm (40 nm).

CONCLUSIONS

In summary, we presented simulations of short-channel ballistic CNT FETs that explain recent experimental results using Pd contacts, and we have reached two important conclusions about the contacts. The first conclusion is that the contacts are strongly modulated by the gate when no bottom metal contact is present, allowing for lower subthreshold swings for short channels and improved scaling behavior. The second conclusion is that field penetration in the small contact cavity is enhanced by the field discontinuity at the dielectric/cavity interface, enhancing the impact of the gate on the contacts. Taken together, our results introduce important design considerations for CNT electronic devices, including more complicated geometries such as CNT FETs with self-aligned gates. In this geometry, the gate electrode does not extend under the source and drain contacts, but fringe fields from the gate could still modulate the contact properties. A more detailed study is necessary to understand the magnitude of the gate-modulation effect in these devices, and could pave the way for high-performance, low-capacitance devices. More generally, our results should also apply to devices made of other nanomaterials such as nanowires and graphene.

METHODS

The potential is obtained from the charge through a 3D solution of Poisson's equation, $\nabla \cdot (\varepsilon \nabla V) = -\rho$, where $\rho$ is the charge density (described below), $V$ is the electrostatic potential, and $\varepsilon$ is the spatially-dependent dielectric constant. We treat the metals in the device as perfect electric conductors by imposing Dirichlet boundary conditions at the edges of the source, drain, and gate electrodes, and we assume Neumann boundary conditions at the left, right, and top edges of the simulation space. Periodic boundary conditions are applied along the $z$-axis (out of the page), and we use a Fourier transform method to accelerate the solution of the 3D Poisson equation. The size of the supercell along this axis is chosen to avoid electrostatic interaction between neighboring CNTs. Poisson's equation is discretized using the finite element method and is solved with a conjugate gradient algorithm, yielding a 3D potential profile, $V(x, y, z)$. The potential along the length of the CNT is then given by $V_{CNT}(x) = \frac{1}{2}[V(x, y_{top}, z_{top}) + V(x, y_{bot}, z_{bot})]$, where $(y_{top}, z_{top})$ are the coordinates on the top of the CNT in Fig. 1a, and $(y_{bot}, z_{bot})$ are the coordinates on the bottom of the CNT. We have found this gives the same results as taking an average potential over the entire circumference of the CNT. For the electrostatic potential calculations the length of the contact region is 10 nm, which ensures that the CNT is in local equilibrium with the metal.

To calculate the charge due to a given potential, we describe the electronic structure of the CNT with a mode-space tight-binding representation,[21,22] assuming a nearest-neighbor coupling of $\gamma = 2.5$ eV.[23,24] The tight-binding Hamiltonian is given by $H_{ll} = eV_{CNT}(x_l) - i\Delta/2$, $H_{2l,2l-1} = H_{2l-1,2l} = 2\gamma\cos(\pi J/M)$, and $H_{2l,2l+1} = H_{2l+1,2l} = \gamma$, where $e$ is the electron charge, $x_l$ is the position of the $l^{th}$ carbon ring, $J$ is the subband index, $M$ is the number of atoms per carbon ring, and $\Delta$ is the CNT-metal coupling strength. The charge density along the CNT is calculated as $\rho_{CNT}(x_l) = e/2\pi \cdot \int \text{Im}\, G^{<}_{ll}(E)\, dE$, where $G^{<}$ is the electron correlation function, determined by applying the tight-binding Hamiltonian to the NEGF formalism.[18,21,22] This 1D charge density is then mapped back to 3D with a Gaussian distribution of the charge around the CNT radius, and by interpolating $\rho_{CNT}$ onto the grid used to solve Poisson's equation. The width of the Gaussian distribution is 0.06 nm.

The ballistic zero-bias conductance is calculated as $G = 4e^2/h \cdot \int T(E)\left[-df(E)/dE\right]dE$, where $h$ is Planck's constant, $f(E)$ the Fermi function, and $T(E)$ is the transmission through the device, determined using the NEGF formalism and the tight-binding Hamiltonian including the self-consistent potential. To simulate contacts of different lengths, we extend the Hamiltonian off the edges of the simulation space so that the total contact length is $L_C$, and we assume a flat potential and a finite value of $\Delta$ in these regions. In the NEGF formalism, these extensions serve as the self-energies due to the leads.

ACKNOWLEDGMENTS

This project is supported by the Laboratory Directed Research and Development program at Sandia National Laboratories, a multiprogram laboratory operated by Sandia Corporation, a Lockheed Martin Co., for the United States Department of Energy under Contract No. DEAC01-94-AL85000.

1. Charlier J.-C.; Blase, X.; Roche, S. Electronic and Transport Properties of Nanotubes. *Rev. Mod. Phys.* **2007**, *79*, 677-732.

2. Avouris, P.; Chen, Z.; Perebeinos, V. Carbon-Based Electronics. *Nat. Nanotechnol.* **2007,** *2*, 605-615.

3. Chen, Z.; Appenzeller, J.; Knoch, J.; Lin, Y.; Avouris, P. The Role of Metal-Nanotube Contact in the Performance of Carbon Nanotube Field-Effect Transistors. *Nano. Lett.* **2005**, *5*, 1497-1502.

4. Léonard, F.; Talin, A. A. Electrical Contacts to One- and Two-Dimensional Nanomaterials. *Nat. Nanotechnol.* **2011**, *6*, 773-783.

5. Léonard, F.; Talin, A. A. Size-Dependent Effects on Electrical Contacts to Nanotubes and Nanowires. *Phys. Rev. Lett.* **2006**, *97*, 026804.

6. Javey, A.; Guo, J.; Wang, Q.; Lundstrom, M.; Dai, H. Ballistic Carbon Nanotube Field-Effect Transistors. *Nature* **2003**, *424*, 654-657.

7. Kim, W.; Javey, A.; Tu, R.; Cao, J.; Wang, Q.; Dai, H. Electrical Contacts to Carbon Nanotubes down to 1 nm in Diameter. *Appl. Phys. Lett*. **2005**, *87*, 173101.

8. Perello, D. J.; Lim, S. C.; Chae, S. J.; Lee, I.; Kim, M. J.; Lee, Y. H.; Yun, M. Thermionic Field Emission Transport in Carbon Nanotube Transistors. *ACS Nano* **2011**, *5*, 1756-1760.

9. Solomon, P. M. Contact Resistance to a One-Dimensional Quasi-Ballistic Nanotube/Wire. *IEEE Electron Device Lett.* **2011**, *32*, 246-248.

10. Nemec, N.; Tománek, D.; Cuniberti, G. Contact Dependence of Carrier Injection in Carbon Nanotubes: An *Ab Initio* Study. *Phys. Rev. Lett.* **2006**, *96*, 076802.

11. Nemec, N.; Tománek, D.; Cuniberti, G. Modeling Extended Contacts for Nanotube and Graphene Devices. *Phys. Rev. B* **2008**, *77*, 125420.

12. Xue, Y.; Ratner, M. A. Scaling Analysis of Schottky Barriers at Metal-Embedded Semiconducting Carbon Nanotube Interfaces. *Phys. Rev. B* **2004**, *69*, 161402.

13. Andriotis, A. N.; Menon, M. Structural and Conducting Properties of Metal Carbon-Nanotube Contacts: Extended Molecule Approximation. *Phys. Rev. B* **2007**, *76*, 045412.

14. Song, X.; Han, X.; Fu, Q.; Xu, J.; Wang, N.; Yu, D.-P. Electrical Transport Measurements of the Side-Contacts and Embedded-End-Contacts of Platinum Leads on the Same Single-Walled Carbon Nanotube. *Nanotechnology* **2009**, *20*, 195202.

15. Franklin, A. D.; Chen, Z. Length Scaling of Carbon Nanotube Transistors. *Nat. Nanotechnol.* **2010**, *5*, 858-862.

16. Franklin, A. D.; Luisier, M.; Han, S.-J.; Tulevski, G.; Breslin, C. M.; Gignac, L.; Lundstrom, M. S.; Haensch, W. Sub-10 nm Carbon Nanotube Transistor. *Nano Lett.* **2012**, *12*, 758-762.

17. Seol, G.; Guo, J. Metal contact to graphene nanoribbon. *Appl. Phys. Lett.* **2012**, *100*, 063108.

18. Datta, S. *Electronic Transport in Mesoscopic Systems;* Cambridge University Press: Cambridge, 1995.

19. Khomyakov, P. A.; Giovannetti, G.; Rusu, P. C.; Brocks, G.; van den Brink, J.; Kelly, P. J. First-Principles Study of the Interaction and Charge Transfer between Graphene and Metals. *Phys. Rev. B* **2009**, *79*, 195425.

20. Greve, D. W. *Field Effect Devices and Applications*; Prentice Hall: Upper Saddle River, New Jersey, 1998.

21. Léonard, F. Quantum Transport in Nanotube Transistors. *Proc. ICCN 2002* **2002**, 298-301.

22. Guo, J.; Datta. S.; Lundstrom, M. S.; Anantram, M. P. Toward Multiscale Modeling of Carbon Nanotube Transistors. *Int. J. Multiscale Com.* **2004**, *2*, 257-276.

23. Mintmire, J. W.; White, C. T. Electronic and Structural Properties of Carbon Nanotubes.

*Carbon* **1995**, *33*, 893-902.

24. Odom, T. W.; Huang, J.-L.; Kim, P.; Lieber, C. M. Atomic Structure and Electronic Properties of Single-Walled Carbon Nanotubes. *Nature* **1998**, *391*, 62-64.